\documentclass[11pt,twoside]{article}
\usepackage{asp2010}

\resetcounters

\makeatletter

\def\ltappeq{\compoundrel<\over\sim}
\def\compoundrel#1\over#2{\mathpalette\compoundreL{{#1}\over{#2}}}
\def\compoundreL#1#2{\compoundREL#1#2}
\def\compoundREL#1#2\over#3{\mathrel
      {\vcenter{\hbox{$\m@th\buildrel{#1#2}\over{#1#3}$}}}}
\makeatother

\begin{document}

\title{The Evolution of Cataclysmic Variables}
\author{Christian Knigge}
\affil{School of Physics \& Astronomy, University of Southampton,
Southampton SO17 1BJ, UK} 

\begin{abstract}
I review our current understanding of the evolution of cataclysmic
variables (CVs). I first
provide a brief introductory ``CV primer'', in 
which I describe the physical structure of CVs, as well as their
astrophysical significance. The main part of the review is divided
into three parts. The first part outlines the theoretical principles of CV
evolution, focusing specifically on the standard ``disrupted magnetic
braking'' model. The second part describes how some of the most fundamental
predictions this model are at last being test
observationally. Finally, the third part describes recent efforts to
actually reconstruct the evolution path of CVs empirically. Some of
these efforts suggest that angular momentum loss below the period gap
must be enhanced relative to the purely gravitational-radiation-driven
losses assumed in the standard model. 
\end{abstract}

\section{Introduction: A CV Primer}

\begin{figure}[t]
  \begin{center}
  \includegraphics[height=.28\textheight]{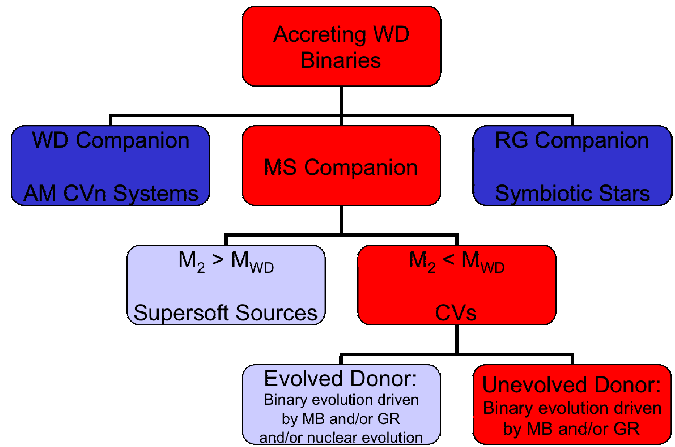}
  \caption{The taxonomy of accreting white dwarf binary systems.}
  \label{fig:zoo}
  \end{center}
\end{figure}

Cataclysmic Variables (CVs) are binary systems containing an accreting
white dwarf (WD). However, this broad class also contains other
types of objects, so the taxonomy may be a little
confusing. Figure~\ref{fig:zoo}
therefore provides a simplified take on how CVs fit into the zoo of
accreting WDs. The key point is that the secondaries in CVs are
stars on or near the main sequence (MS). In fact, for the purpose of
this review, I will focus almost exclusively on 
CVs in which the secondaries have undergone no significant nuclear
evolution at the onset of mass transfer. These dominate the CV
population below $P_{orb} \simeq 4-5$~hr
\citep{1998A&A...339..518B,2003MNRAS.340.1214P,2006MNRAS.373..484K}.
In fact, I will further concentrate on {\em non-magnetic} CVs,
i.e. systems in which the magnetic field of the primary WD is too weak
to affect the accretion flow. This restriction matters, since the much
rarer magnetic CVs may evolve differently from non-magnetic ones
\citep[e.g.][]{1995ASSL..205..315W,2009ApJ...693.1007T}.

\begin{figure}[t]
  \begin{center}
  \includegraphics[height=.25\textheight]{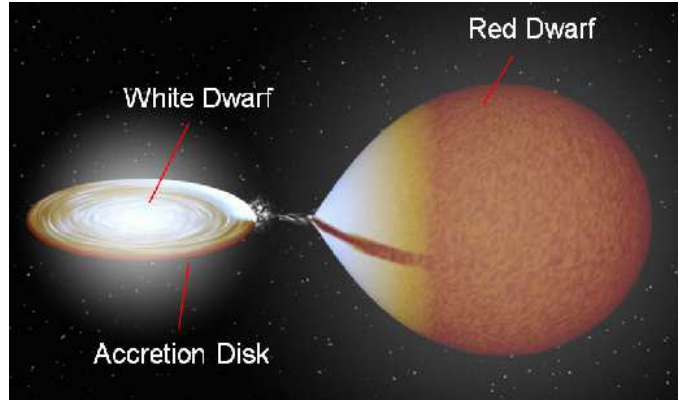}
  \caption{The physical structure of CVs. The key components 
are a roughly MS Roche-lobe-filling secondary star, an accreting 
WD primary, and an accretion disk surrounding the WD. This figure was
created by Rob Hynes using his BinSim binary visualization code (see http://www.phys.lsu.edu/~rih/binsim).} 
  \label{fig:hynes}
  \end{center}
\end{figure}

The physical structure of non-magnetic CVs is illustrated in
Figure~\ref{fig:hynes}. CVs are {\em close} binary systems, 
with binary separations $a_{bin} \sim R_{\odot}$ and orbital periods
$80~{\rm min} \ltappeq P_{orb} 6~{\rm hr}$. As a result, the
secondary undergoes Roche-lobe overflow and loses mass through the
inner Lagrangian point. Since this material has excess angular
momentum, it is then transported towards the WD via an accretion
disk. 

\begin{figure}[t]
  \begin{center}
  \includegraphics[height=.35\textheight]{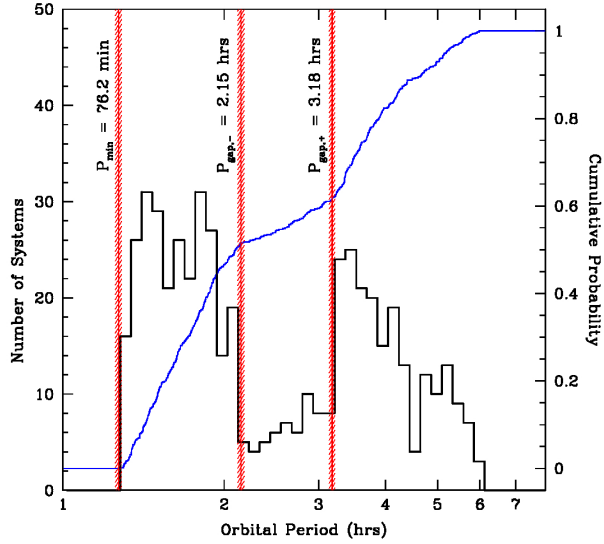}
  \caption{Differential and cumulative orbital period distribution of
  CVs, based on data taken from Edition 7.6 of the Ritter \& Kolb
  catalogue \citep{2003A&A...404..301R}. Estimated values for the minimum period and the
  period gap edges are shown as vertical lines. The shaded regions
  around them indicate our estimate of the errors on these
  values. Figure reproduced from \citet{2006MNRAS.373..484K}.}
  \label{fig:pdist}
  \end{center}
\end{figure}

The mass transfer process in CVs appears to be relatively stable on
long time scales. This requires both that the mass ratio $q = M_2/M_1
\ltappeq 1$ and also the existence of an angular momentum loss (AML)
process that continually shrinks the system and thus keeps the Roche
lobe in touch with the secondary star. The orbital period therefore 
initially decreases as a CV evolves, making the period distribution a
powerful tracer of CV evolution. Figure~\ref{fig:pdist} shows this period
distribution for CVs. The two most obvious features in this
distribution are (i) the famous ``period gap'' between $P_{orb} \simeq
2~{\rm hr}$ and $P_{orb} \simeq 3~{\rm hr}$; (ii) a sharp cut-off
around a minimum period of $P_{min} \simeq 80~{\rm min}$. 

In what has become the ``standard model'' for CV
evolution, mass transfer {\em above} the gap is driven primarily by
AMLs associated with a weak,
magnetized stellar wind from the secondary (``magnetic braking''
[MB]), while mass
transfer {\em below} the gap is driven solely by gravitational
radiation (GR). As we shall see in more detail
below, the period gap then arises as a consequence of the cessation of 
MB at $P_{orb} \simeq 3~{\rm hr}$. It is usually assumed that this
cessation is associated with the transition of the star from a partly
radiative structure to a fully convective one. The period minimum is
also associated with a change in the structure of the donor. Roughly
speaking (see below), $P_{min}$ marks the transition of the donor from a
star to a sub-stellar object. Since the radius of a brown dwarf {\em
increases} in response to mass loss, this transition must also lead to
a change in the direction of orbital period evolution. Systems that
have already evolved {\em beyond} $P_{min}$ are often referred to as
``period bouncers''.

\begin{figure}[t]
  \begin{center}
  \includegraphics[height=.45\textheight]{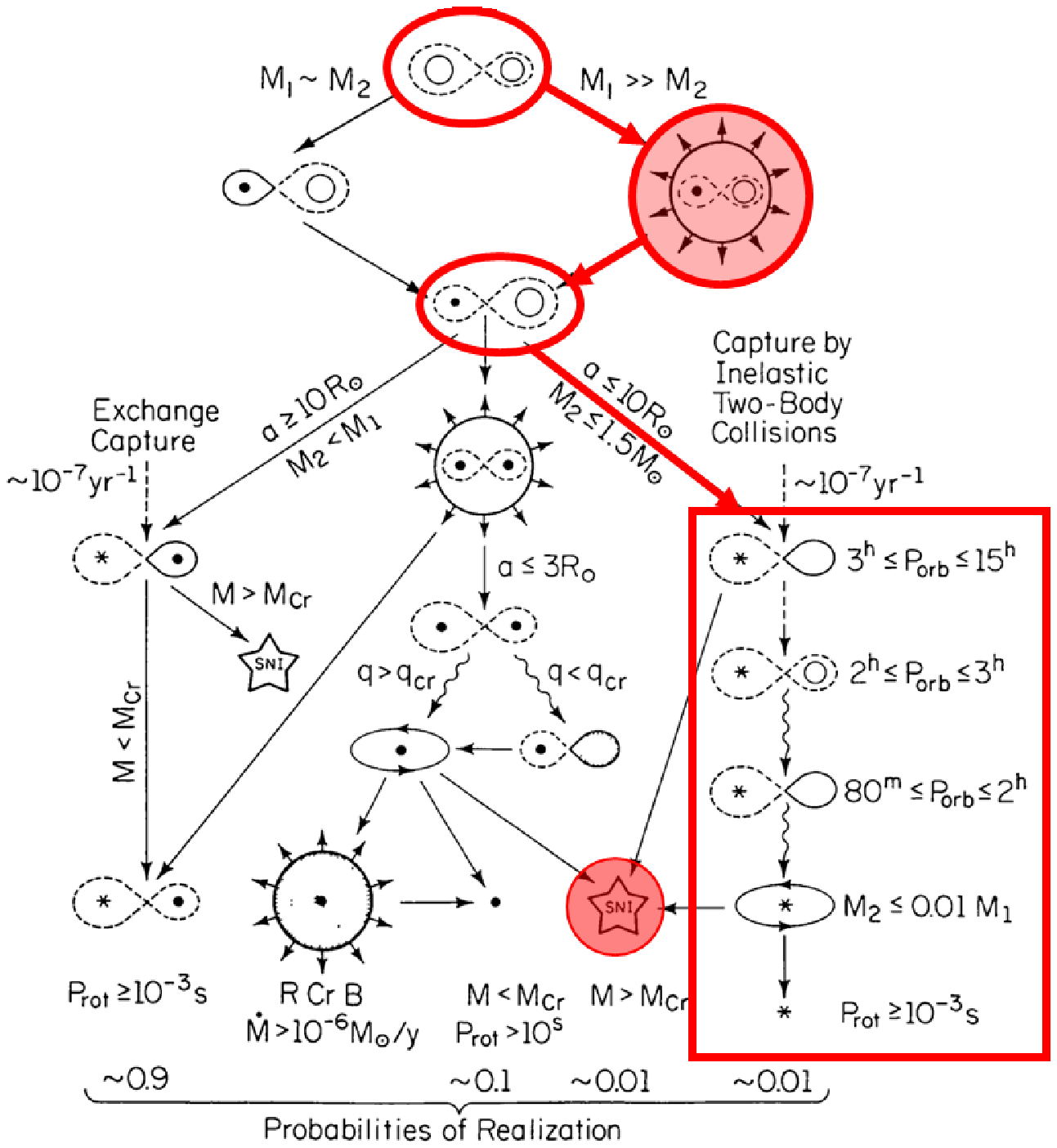}
  \caption{Basic scenarios for binary star evolution after
    \citet{1991ApJS...76...55I} and \citet{1984ApJS...54..335I}. Rings are 
    unevolved stars; filled circles are elctron-degenerate helium,
    carbon-oxygen, or oxygen-neon cores in giants; six-pointed stars
    are white dwarfs or neutron stars. Wavy lines mark transitions
    driven by the radiation of gravitational waves and
    counter-clockwise rotating ellipss are heavy disks. Open stars
    represent Type Ia supernova explositions. Roche lobes are shown by
    dashed loops (when not filled) or by solid loops (when
    filled). The probability of realization of different final
    products is indicated at the bottom of the figure. The
    evolutionary channel marked in red is that expected to produce
    most CVs. The CV phase itself is marked by the red box. Figure
    adapted from \citet{1991ApJS...76...55I}.}
  \label{fig:iben}
  \end{center}
\end{figure}

Understanding the evolution of CVs -- in the first instance by testing
this standard model -- is astrophysically important. This is
illustrated in Figure~\ref{fig:iben}, which places CVs in the context of other
close binary systems. First, even though many binary stars interact
with each other at some stage of their lives, CVs provide a rare
opportunity to observe {\em long-lived stable} mass transfer in
action. Second, the physical processes that are relevant to CV
evolution -- not just MB and GR, but also the infamous ``common
envelope'' (CE) phase that brings these systems into (or close to)
contact -- are also key to many other types of binary systems. 
Third, some CVs and their 
relatives (such as the supersoft sources, see Figure~\ref{fig:zoo}) are expected
to be Type Ia supernova progenitors. Fourth and finally, it is
becomingly increasingly clear that most aspects of the accretion
process in neutron star and black hole binary systems -- including
variability \citep[e.g.][]{2005AIPC..797..277W}, accretion disk winds
\citep[e.g.]{1982ApJ...260..716C,2002ApJ...579..725L}, and jets
\cite{2008Sci...320.1318K} -- have 
direct counterparts in CVs. Since CVs are relatively numerous, nearby,  
bright and characterized by observationally ``convenient'' orbital 
time-scales, this makes them extremely useful as laboratories for the
study of accretion onto compact objects more generally.

\section{Principles of CV Evolution: Theory}

The evolution of CVs is closely connected to -- and in some sense
controlled by -- the properties of their secondary stars. Indeed, as 
noted above, mass transfer above the period gap is driven by MB (a process 
associated with the secondary), and both the period gap and the period
minimum are thought to mark structural changes in the secondary. In
order to understand CV evolution, we must therefore understand the
properties of their donor stars. Much of the following is therefore
reproduced from \citet{kyoto}, which reviews our current understanding
of CV secondaries. 

\subsection{Fundamentals}

The radius of a Roche-lobe-filling star depends only the binary
separation, $a$, and the mass ratio, $q = M_2/M_1$. A particularly
convenient approximation for the Roche-lobe radius is
\citep{1971ARA&A...9..183P}
\begin{equation}
\frac{R_L}{a} = \frac{2}{3^{4/3}} \left[\frac{q}{1+q}\right]^{1/3},
\end{equation}
which can be combined with Kepler's third law
\begin{equation}
P_{orb}^2 = \frac{4\pi^2a^3}{G(M_1+M_2)}
\end{equation}
to yield the well-known {\em period-density relation} for
Roche-lobe-filling stars with $R_2 = R_L$ 
\begin{equation}
\left< \rho_2 \right>  = \frac{M_2}{(4\pi/3)R^3_2} \simeq 100 G^{-1}P_{orb}^{-2}
.
\label{eq:pden}
\end{equation}
Let us assume for the moment that CV donors are indeed mostly low-mass,
near-MS stars. We then expect that their mass-radius relationship will
be roughly
\begin{equation}
R_2/R_{\odot} = f (M_2/M_{\odot})^{\alpha}
\label{eq:pow_mr}
\end{equation}
with $f \simeq \alpha \simeq 1$. Combining this with the
period-density relation immediately gives approximate
mass-period and radius-period relations for CV donors 
\begin{equation}
M_2/M_{\odot} = M_2/M_{\odot} \simeq 0.1 P_{orb,hr},
\end{equation}
where $P_{orb,hr}$ is the orbital period in units of hours. This shows
that the period gap between 2~hrs and 3~hrs corresponds to $M_2 \simeq
0.2 - 0.3 M_{\odot}$, which is, in fact, where the secondary is
expected to change structure rom partly
radiative to fully convective.

\subsection{Are CV Donors on the Main Sequence?}
\label{sec:ms}

How good is the assumption that CV donors are nearly ordinary MS stars?
The answer depends on the competition between the mass-loss time
scale, on which the ongoing mass transfer reduces the donor mass, 
\begin{equation}
\tau_{\dot{M}_2} \simeq \frac{M_2}{\dot{M}_2}.
\label{eq:tau_mdot}
\end{equation}
and the thermal time scale, on which the donor can correct deviations from
thermal equilibrium (TE),
\begin{equation}
\tau_{th} \simeq \frac{GM_2^2}{L_2R_2} \simeq 10^8 (M_2/M_{\odot})^{-3/2} {\rm yrs}.
\label{eq:tau_kh}
\end{equation}
If mass loss is slow, in the sense that $\tau_{\dot{M}_2} >>
\tau_{th}$, the donor always has time to adjust itself to attain the
appropriate TE structure for its current mass. It
therefore remains on the MS, with a mass-radius index $\alpha \simeq
1$, and is essentially indistuinguishable from an isolated MS star. 
Conversely, if mass loss is fast, i.e. $\tau_{\dot{M}_2} <<
\tau_{th}$, the donor cannot adjust its structure quickly enough to
remain in TE. Instead, the mass loss is effectively adiabatic. The
response of low-mass stars with at least a substantial convective
envelope to such mass loss is to {\em expand}, with $\alpha \simeq
-1/3$. 

Which of these limits is appropriate for CVs? Neither, as it turns
out. Let us take some typical parameters suggested by the standard model
for CVs above and below the gap, say $M_2 \simeq 0.4$ with $\dot{M}_2 \simeq
1\times10^{-9}$ and $M_2 \simeq 0.1$ with  $\dot{M}_2 \simeq
3\times10^{-11}$, respectively. Plugging these values into
Equations~\ref{eq:tau_mdot} and \ref{eq:tau_kh}, we find $\tau_{\dot{M}_2}
\simeq \tau_{th} \simeq 4 \times 10^{8}$~yrs above the gap and
$\tau_{\dot{M}_2} \simeq \tau_{th} \simeq 3 \times 10^{9}$~yrs
below. {\em Thus the thermal and mass-loss time scales are comparable
for CV donors, both above and below the period gap.}

What this means is that the donor cannot {\em quite} shrink fast
enough to keep up with the rate at which mass is 
removed from its surface. It is therefore driven slightly
out of thermal equilibrium and becomes somewhat oversized for its
mass. It is this slight deviation from TE that ultimately explains
both the period gap and the period minimum.

\subsection{The Origin of the Period Gap}

Let us take as given, for the moment, that the period gap is
``somehow'' associated with a sudden cessation of (or at least 
reduction in) MB a $P_{orb} \simeq 3$~hrs. Why should this produce a
period gap in the CV population?
 
Recall that the donor star is slightly out of thermal equilibrium --
i.e. slightly bloated -- as it encounters the upper edge of the period
gap. Now, since mass transfer in CVs is driven entirely by AML, a
sudden reduction in AML will also result in a sudden reduction in the
mass-loss rate the donor experiences. This lower mass-loss
rate cannot sustain the same degree of thermal disequilibrium
and inflation in the secondary star. The donor therefore responds to 
this change by shrinking closer to its thermal equilibrium
radius. This results in a loss of contact with the Roche lobe.

The upper edge of the gap thus marks a cessation of mass transfer in
CVs. According to the standard model, CVs then evolve through the period 
gap as detached systems. During this detached phase, the binary orbit
and Roche lobe continue to shrink, since AML due to GR
continues. However, since the thermal relaxation of the
donor in this phase is faster than the shrinkage of the Roche lobe,
the donor manages to relax all the way back to its TE radius. 
The bottom edge of the period gap thus corresponds to the location
where the Roche lobe radius catches up once again to the TE radius of
the donor. At this point, mass transfer restarts, and the system
emerges from the gap as an active CV once again.

How bloated must CV donors be to account for the observed size of the
period gap? Since there is no mass transfer {\em in} the gap, the
donor mass just above and below the gap must be the same,
$M_2(P_{gap,+}) = M_2(P_{gap,-})$. From the period-density relation
(Equation~\ref{eq:pden}), we then get 
\begin{equation}
\frac{R_2(P_{gap,+})}{R_2(P_{gap,-})} =
\left[\frac{P_{gap,+}}{P_{gap,-}} \right]^{2/3} \simeq
\left[\frac{3}{2}\right]^{2/3} \simeq 1.3.
\end{equation}
We also know that the donor at the bottom edge is in or near
equilibrium, so that {\em donors at the upper edge of the period gap
must be oversized by $\simeq 30\%$ relative to equal-mass, isolated
MS stars.}

\subsection{The Origin of the Period Minimum}

The period minimum is also closely connected to the properties of the
donor stars. If we combine the period-density relation 
(Equation~\ref{eq:pden}) with the simple power-law approximation to the
donor mass-radius relation (Equation~\ref{eq:pow_mr}), we find
\begin{equation}
P_{orb}^{-2} \propto M_2^{1-3\alpha}.
\end{equation}
Differentiating this logarithmically yields a simple expression for the
orbital period derivative, i.e.
\begin{equation}
\frac{\dot{P}_{orb}}{P_{orb}} = \frac{3\alpha-1}{2} \frac{\dot{M}_2}{M_2}.
\label{eq:pdot}
\end{equation}
Since the period minimum must correspond to $\dot{P}_{orb} = 0$,
Equation~\ref{eq:pdot} tells us that $P_{min}$ occurs when the donor
has been driven so far out of thermal equilibrium that its mass-radius
index along the evolution track has been reduced from its near-MS
value of $\alpha \simeq 1$ to $\alpha = 1/3$. So, as already noted
above, $P_{min}$ does not necessarily have to coincide with the
orbital period at which the donor mass reaches the Hydrogen-burning
limit, $M_H$. In fact, recall
that we noted in Section~\ref{sec:ms} that, for 
any donor with at least a substantial convective envelope, the
mass-radius index in the limit of {\em fast} (adiabatic)
mass-transfer is $\alpha \simeq -1/3$. Thus the period evolution of a
CV can in principle be made to turn around at {\em any} donor mass,
provided only that mass loss becomes sufficiently fast compared to the
donor's thermal time scale. The significance of $M_H$ in this context
is that period bounce becomes inevitable when the donor reaches
this limit. This is because sub-stellar objects are out of TE by
definition and respond even to slow mass loss by increasing in radius,
i.e. $\alpha \leq 0$. In practice, $P_{min}$ does, in fact, correspond
roughly to $M_2 \simeq M_H$.  

\subsection{Magnetic Braking}

\begin{figure}[t]
  \begin{center}
  \includegraphics[height=.40\textheight,angle=-90]{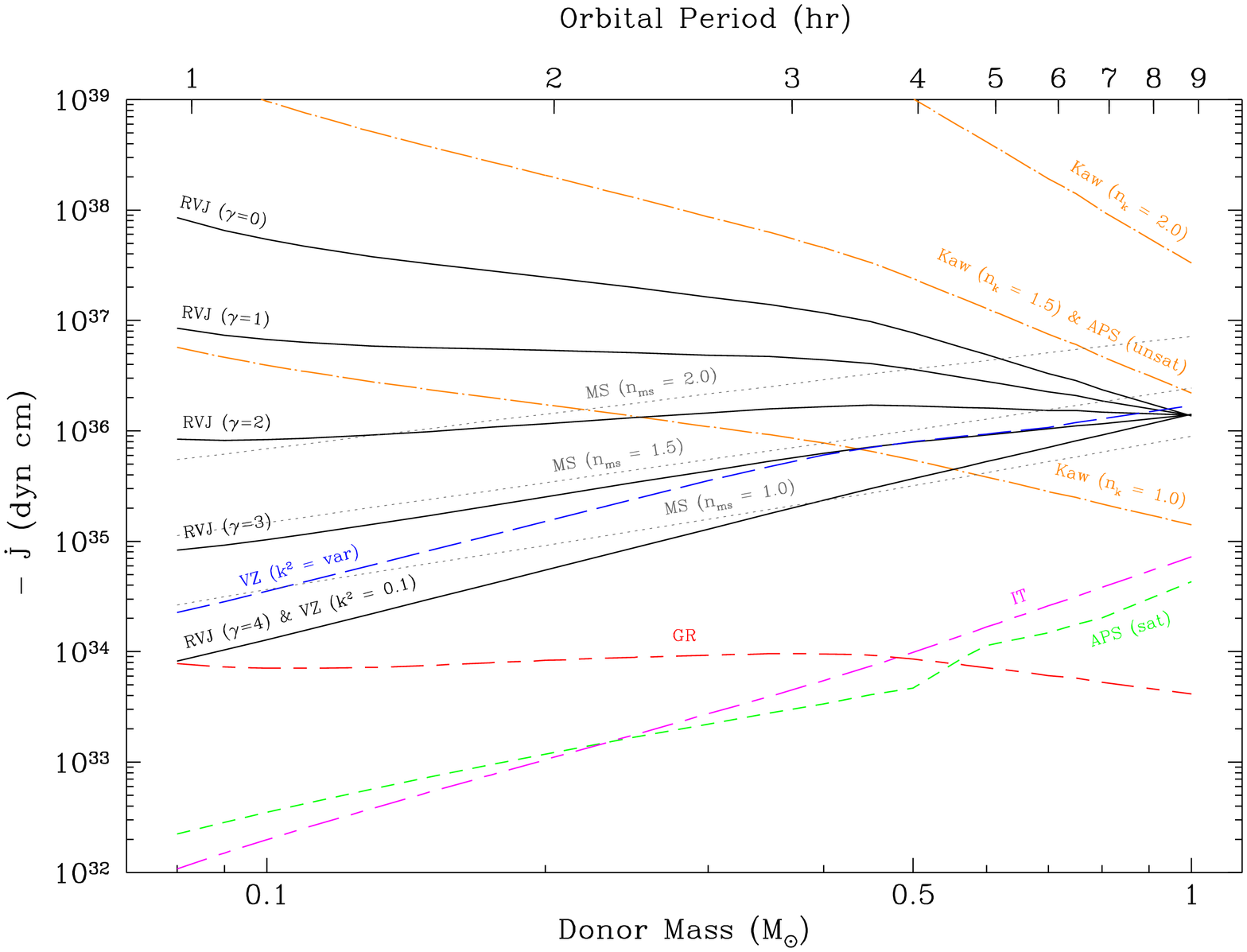}
  \caption{A comparison of several widely used angular momentum loss
recipes. The red long-dash-short-dash line corresponds to GR-driven
AML; the blue long-dashed lines the classic
\citet{1981A&A...100L...7V} prescription; the solid black lines mark
the \citet{1983ApJ...275..713R} prescription; the orange dash-dotted
line mark the \citet{1988ApJ...333..236K} recipe; the gray dotted
lines mark the \citet{1987MNRAS.226...57M} law; the green short-dashed
line shows the saturated model suggested by
\citet{2003ApJ...582..358A}; long-dashed magenta line shows the 
saturated AML law suggested by \citet{2003ApJ...599..516I}. For
several recipes, several versions with different parameters are
shown. The main thing to take away from this figure is the enormous
range in shapes and strengths, even for conceptually similar AML
prescriptions. Figure reproduced from \citet{2011ApJS..194...28K},
which also provides more details on the various AML recipes.}
  \end{center}
\label{fig:mb1}
\end{figure}

As noted above, stable mass transfer in CVs {\em requires} AML from
the system, so understanding the process(es) by which this happens is
key to understanding CV evolution. There is no great mystery about GR,
of course, which (in the standard model) is thought to dominate below
the period gap. However, how well do we really understand MB, the
process expected to dominate above the gap, whose cessation is thought
to {\em cause} the gap?

Figure~\ref{fig:mb1}, taken from \citet{2011ApJS..194...28K}, is an
attempt to answer this question. For this figure, we compiled a
variety of widely used MB recipes from the literature and compared the
AML rates they predict for CV donors. In order to translate the
predictions into orbital period space, we made the ``marginal
contact'' approximation, i.e. we assumed CV secondaries follow the
standard MS mass-radius relation and then used the period-density
relation to estimate $P_{orb}$. For reference, we also extrapolated
the predicted AML recipes into the fully convective regime.

A detailed discussion of the various MB prescriptions shown in
Figure~\ref{fig:mb1} is given in \citet{2011ApJS..194...28K}. 
However, the basic message is
clear: there are huge differences between different recipes, not
just in the rates they predict at fixed $P_{orb}$, but even in the
{\em shape} of the AML rates they predict as a function of
$P_{orb}$. To make matters worse, it has long been known that fully
convective single stars manage to sustain significant magnetic fields,
the key physical requirement for MB. 

\begin{figure}[t]
  \begin{center}
  \includegraphics[height=.25\textheight]{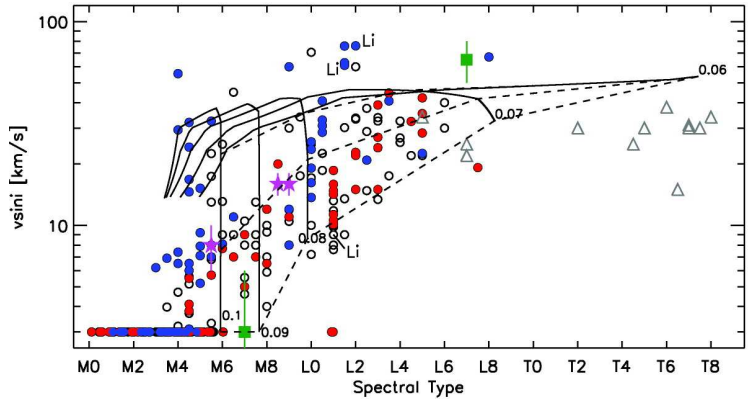}
  \caption{Measurements of $v \sin{i}$ for single stars. 
Filled blue circles are probably young, filled red circles old. The
meaning of the remaining symbols and colours is not relevant to the
discussion here. The solid lines mark evolutionary tracks for objects
of 0.1, 0.09, 0.08, and 0.07 Modot; dashed lines mark ages of 2, 5,
and 10 Gyr (from upper left to lower right). Figure adapted from
\citet{2008ApJ...684.1390R}.}
  \end{center}
\label{fig:reiners}
\end{figure}

Does this mean that the whole idea of disrupted MB as a key driver of CV
is without any basis? Not quite. Figure~\ref{fig:reiners} shows the
rotation velocities of single MS stars as a function of spectral
type\cite{2008ApJ...684.1390R}. What is striking here is that all
stars earlier than about M5 
are extremely slow rotators, whereas stars later than about M5 are
characterized by a wide range of higher rotation rates. Now this
spectral type -- M5 -- actually corresponds roughly to the dividing
line between fully convective and partly radiative stars. So
there is actually some empirical evidence that mass-losing stars may
experience a reduction (if not necessarily a complete cessation) in MB
as they cross this dividing line 
\citep[see also][]{2010A&A...513L...7S}. I think an important implication of
Figure~\ref{fig:reiners} is that researchers studying MB in single and
binary stars have a lot to learn from each other.

\section{Principles of CV Evolution: Observations}

The standard ``disrupted magnetic braking'' model outlined above has
been the cornerstone of CV evolution theory for almost three
decades
\citep{robinson1981,1982ApJ...254..616R,1983ApJ...275..713R,1983A&A...124..267S}.  
It provides reasonable explanations for the existence of the  
period gap and the period minimum, but then this is what it was {\em
designed} to do. Perhaps surprisingly, direct observational tests of
other fundamental predictions it makes have only become possible
over the last few years. Below, I will take a look at three key tests
that have been carried out in this context. Here again, I will
reproduce material from \citet{2010AIPC.1314..171K}, which provides a
broader review of recent observational breakthroughs in CV research. 

\subsection{Disrupted Angular Momentum Loss at the Period Gap}

It is remarkably difficult to test the idea that the gap is caused
specifically by a disruption of AML -- as 
opposed to, for example, the presence of distinct populations above and
below the gap \citep[e.g.][]{2003ApJ...582..358A}. However, there is one key
prediction of the 
model that can, in principle, be tested: if the standard model is correct,
donors just above and below the gap should have identical masses, but
different radii. After all, the donors above the gap have been
significantly inflated by mass loss, while CVs below the gap
have just emerged from a detached phase with their donors in thermal
equilibrium. 

\begin{figure}[t]
  \begin{center}
  \includegraphics[height=.28\textheight]{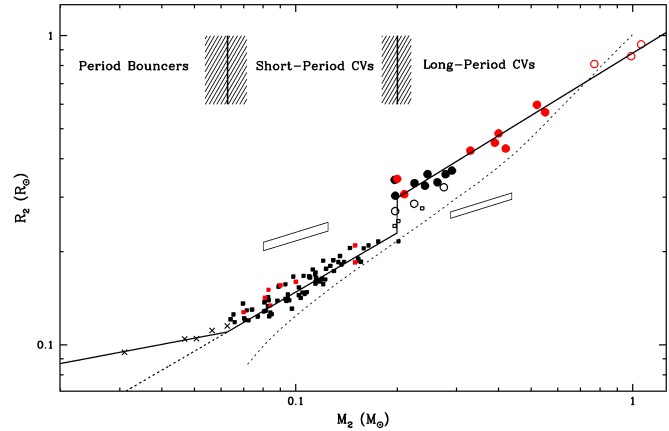}
  \end{center}
  \caption{The mass-radius relation of CV donor stars, based on the
  data presented in \citet{2005PASP..117.1204P}. Superhumpers are shown
  in black, eclipsers in red. Filled 
  squares (circles) correspond to short-period (long-period) CVs,
  crosses to likely period bouncers. The 
  parallelograms illustrate typical errors. 
  Open symbols correspond to systems in the period gap or likely
  evolved systems. The solid lines show the optimal broken power-law 
  fit to the data. The
  dotted line is a theoretical mass-radius relation for MS stars
  \citep{1998A&A...337..403B}. Figure
  reproduced and adapted from \citet{2006MNRAS.373..484K}.} 
\label{fig:mr}
\end{figure}

In 2005, Joe Patterson showed for the first time that this fundamental
prediction is correct \citep{2005PASP..117.1204P}. Over almost two
decades of painstaking work, he and his ``Center for Backyard Astronomy''
collaborators collected a vast amount of observational data on ``superhumps''
in CVs and showed that these observations can be calibrated to yield mass 
ratios for these systems. These mass ratios, in turn, can be used to
obtain estimates of the corresponding donor masses and radii. 
He then combined these with similar estimates obtained for eclipsing
CVs (such estimates are more precise, but available for 
far fewer systems) and put together the mass-radius relationship for
CV donor 
stars shown in Figure~\ref{fig:mr}.\footnote{Actually, the figure here
is from \citet{2006MNRAS.373..484K}, but the data are based entirely on
Patterson's compilation in \citet{2005PASP..117.1204P}.}

The main result is immediately apparent: there is a clear
discontinuity in donor radii at $M_2 \simeq 0.2 
M_{\odot}$ that also cleanly separates long-period from short-period
systems. In fact, donors in systems just below the period gap have
radii consistent with ordinary MS stars of equal mass, while donors
just above the gap have radii that are inflated by $\simeq 30\%$. All
of these findings are exactly in line with the basic predictions of
the disrupted MB model.

Before moving on, it is worth emphasizing that Figure~\ref{fig:mr} alone
cannot tell us the exact nature of the disruption in AML responsible
for the period gap. In particular, {\em any} significant reduction of
AML at $P \simeq 3$~hrs will produce a period gap and a 
discontinuity in the donor mass-radius relationship. Without further
modeling, the data cannot tell us if the AML above the gap has the 
strength expected for MB, nor if MB ceased completely or was merely 
somewhat suppressed at the upper gap edge. However, Figure~\ref{fig:mr}
is extremely strong evidence for the basic idea of a disruption in 
AML at the upper gap edege. 

\begin{figure}[t]
  \begin{center}
  \includegraphics[height=.35\textheight]{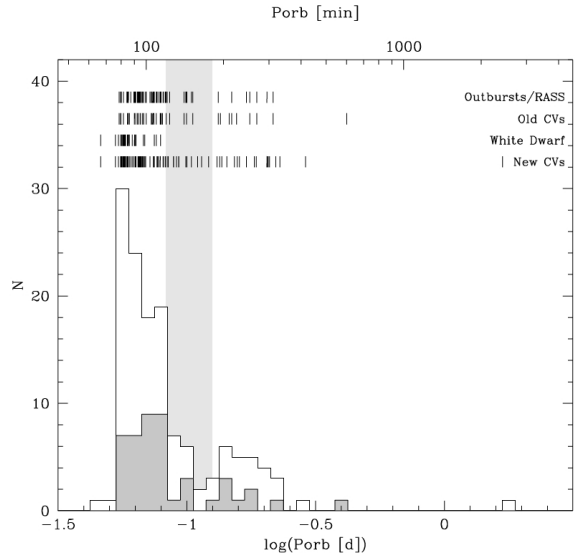}
  \end{center}
  \caption{The period
  distribution of SDSS CVs, divided into 45 previously known
  systems (old SDSS CVs, grey) and 92 newly identified CVs (new SDSS
  CVs, white). Superimposed are tick marks indicating the individual
  orbital periods of the old and new SDSS CVs, those of SDSS CVs
  showing outbursts, those of SDSS CVs detected in the Rosat 
  All-Sky Survey, and those of SDSS CVs which reveal
  the WD in their optical spectra. Figure  adapted and reproduced 
  from \citet{2009MNRAS.397.2170G}.}
\label{fig:spike}
\end{figure}

\subsection{The Period Spike: the Reversal of the Direction of
Period Evolution at $P_{min}$}

Another long-standing prediction of the basic evolution scenario
for CVs is that there should be a ``period spike'' at the 
minimum period \citep[e.g.][]{1999MNRAS.309.1034K}. More specifically,
the orbital period 
distribution of any sufficiently deep sample should show at least a
local maximum near $P_{min}$. This prediction is easy to understand:
the number of CVs we should expect to find in any period interval is
proportional to the time it takes a CV to cross this
interval, $N(P) \propto \dot{P}^{-1}$. But $\dot{P}(P_{min}) = 0$, 
so the period interval including $P_{min}$ should contain an unusually 
large number of systems. This is a critical prediction, since it
follows directly from the idea that $P_{min}$ marks a change in the
direction of evolution for CVs.

Until recently, no CV sample or catalogue showed any sign of the
expected period spike (e.g. Figure~\ref{fig:pdist}). 
However, CVs near $P_{min}$ are very faint, so it was recognized
that this could just be due to a lack of depth in these
samples\citep{2003MNRAS.340..623B}

Enter the sample of $\simeq 200$ CVs constructed by Paula Szkody and
collaborators from the Sloan Digital Sky Survey
\citep[SDSS;][]{szkody1,szkody2,szkody3,szkody4,szkody5,szkody6,2009AJ....137.4011S}. 
This sample 
has a much deeper effective magnitude limit than previous ones and is
therefore much more sensitive to the very faint CVs near and beyond
$P_{min}$. However, in order to test for the existence of a period
spike in this sample, precise orbital periods are needed. These were
obtained via a long-term observational effort led by Boris
G\"{a}nsicke\citep{2009MNRAS.397.2170G}. Figure~\ref{fig:spike} shows the
resulting period distribution for the SDSS CVs. The period spike near
$P_{min}$ is clearly visible.

Now, the existence of the period spike does not necessarily imply that
the standard model is quantitatively correct. In particular, it does not
mean that 
AML below the gap must be driven solely by GR. In fact, the location
of the spike at $P_{min} \simeq 82$~min is quite far from the 
prediction of the standard model \citep[$P_{min} \simeq 65
-70$~min; e.g.][]{1993A&A...271..149K}. Stronger-than-GR AML below the gap 
may be required to 
reconcile this discrepancy between theory and observations. However, 
the discovery of the period spike in the SDSS sample does provide
convincing evidence for the fundamental prediction that CVs actually
undergo a period {\em bounce} at $P_{min}$. 

\subsection{The Existence of CVs with Brown Dwarf Secondaries}

A third key prediction of the standard model of CV evolution is that
most CVs should already have evolved past the period minimum, i.e. they
should be ``post-period-minimum systems'' or ``period
bouncers''. In fact, the standard model predicts that about 70\% of
present day CVs should be period bouncers, with all of these
possessing sub-stellar donor stars \citep[e.g.][]{1993A&A...271..149K}. It was
therefore quite disconcerting that, until recently, only a handful of
{\em candidate} period bouncers were known. In particular, there was
not even one CV with a well-determined donor mass below the
Hydrogen-burning limit.

This situation has also changed for the better thanks to the SDSS CV
sample. Crucially, this sample included several new {\em eclipsing}
candidate period bouncers, for which component masses could be
determined geometrically by careful modelling of high-quality eclipse
observations.

\begin{figure}[t]
  \begin{center}
  \includegraphics[height=.15\textheight]{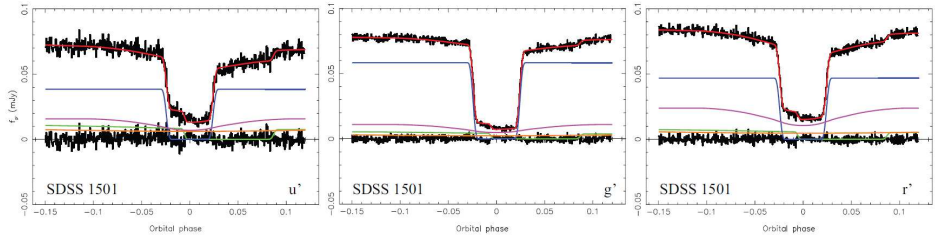}
  \end{center}
  \caption{Model fits to the phase-folded u$^\prime$, g$^\prime$ and
  r$^\prime$ light curves of 
  SDSS J1501. The data (black) are shown with the fit (red) overlaid
  and the residuals plotted below (black). Below are the separate
  light curves of the WD (blue), bright spot (green), accretion disc
  (purple) and the secondary star (orange). Figure 
  reproduced and adapted from \citet{2008MNRAS.388.1582L}.}
\label{fig:eclipse}
\end{figure}

Such eclipse analyses have been carried out by Stuart Littlefair and
collaborators
\citep{littlefair2006,2008MNRAS.388.1582L,2011MNRAS.415.2025S}
have so far yielded three significantly sub-stellar donor mass
estimates, although one of these turns out to be a halo CV (which is
interesting in its own right; Uthas et al. 2011). An
example of a light curve and model fit for one of these systems --
SDSS J1501, whose donor has a mass of $M_2 = 0.053 \pm 0.003
M_{\odot}$ -- is shown in Figure~\ref{fig:eclipse}.

The definitive detection of CVs with sub-stellar donors does not prove
that the standard model is correct -- it is still far from clear, for
example, whether there are enough of these systems in the Galaxy to be
consistent with theoretical predictions. However, it does confirm the
fundamental idea that (at least some) systems survive the
stellar-to-substellar transition of their secondaries, while remaining
active, mass-transferring CVs. 

\begin{figure}[t]
  \begin{center}
  \includegraphics[height=.24\textheight]{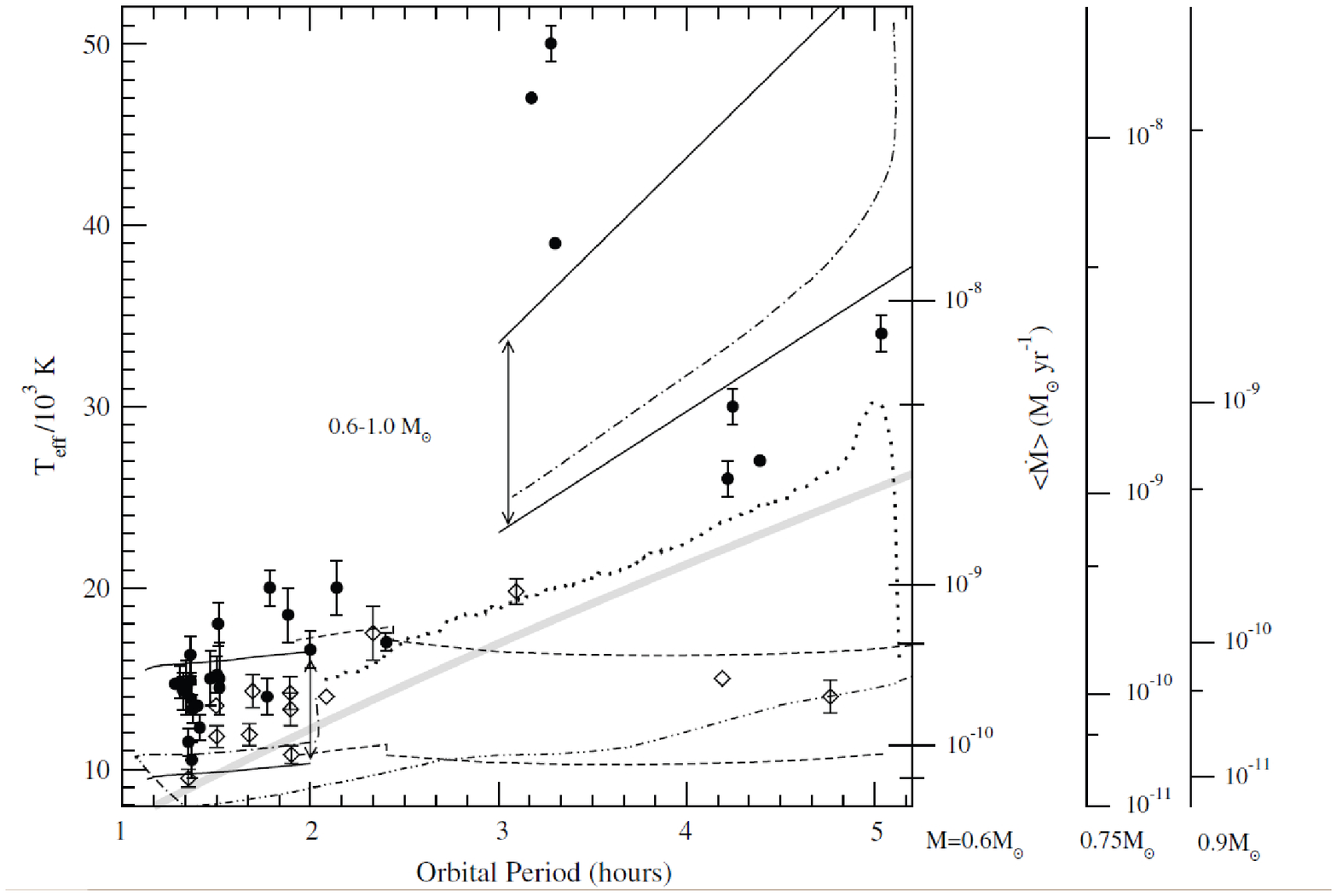}
  \includegraphics[height=.29\textheight]{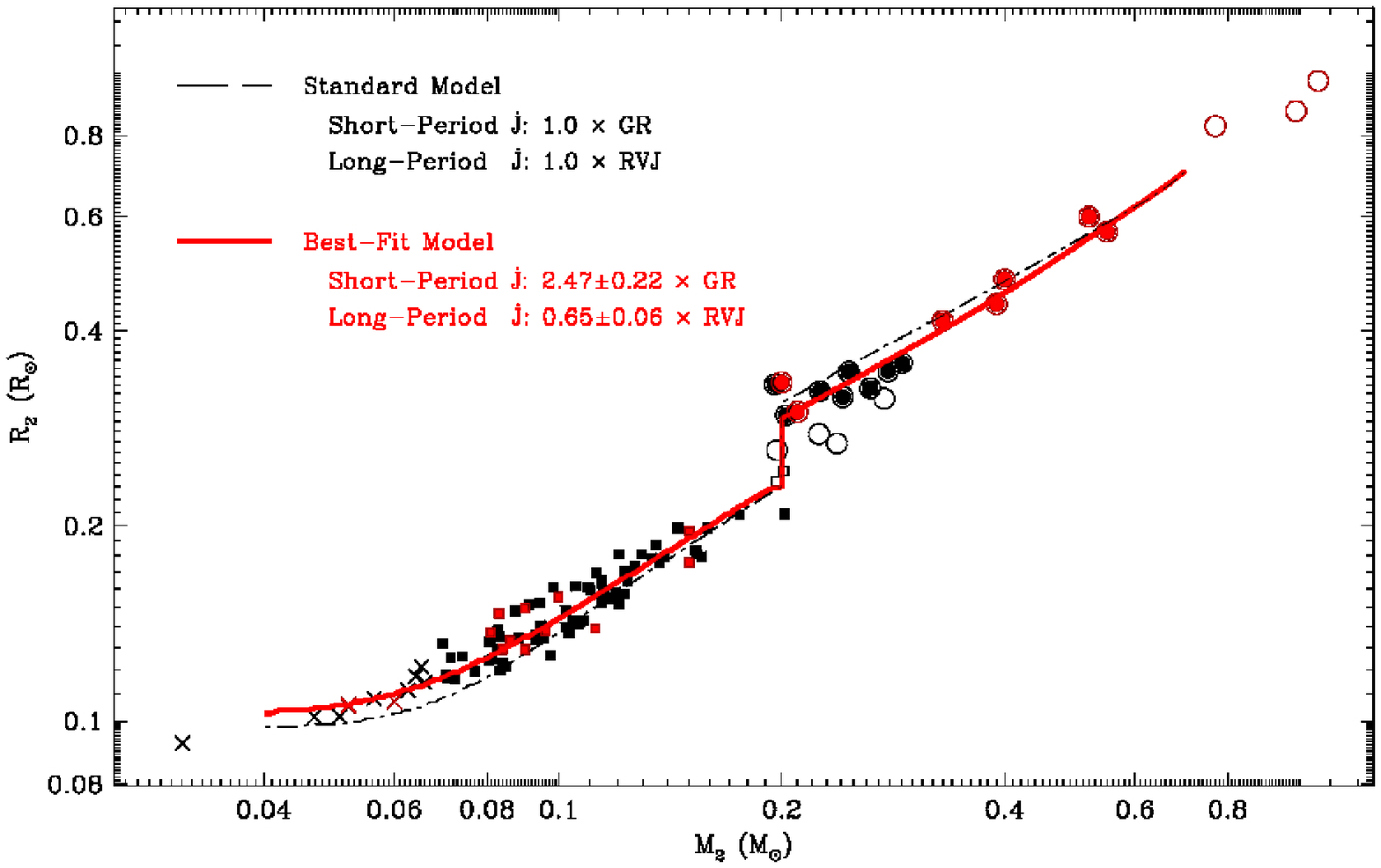}
  \end{center}
  \caption{\scriptsize{\em Top Panel:} Reliable $T_{eff}$ measurements for
  CVs. An approximate mapping to
  $\dot{M}$ is shown on the right vertical scale
  assuming $M_{WD} = 0.75 M_{\odot}$, $0.6 M_{\odot}$, or $0.9 M_{\odot}$. 
  Several sets of predicted temperatures are also
  indicated: an empirical relation \citep[][thick gray
  line]{1984ApJS...54..443P}, traditional MB 
  \citep[][dot-dashed line and between solid lines]{2001ApJ...550..897H}, 
  two versions ofreduced MB due to \citet[][dot-dot-dash
  line]{2003ApJ...582..358A} and \citet[][dotted
  line]{2003ApJ...599..516I}, pure GR (between dashed lines). 
  Figure reproduced from \citet{2009ApJ...693.1007T}.
  {\em Bottom Panel:} Model fits to the
  observed CV donor mass-radius. The thin 
  dashed line is the relationship predicted by the standard
  model, the thick solid line shows the optimal fit achieved by 
  varying the strength of AML above and below the gap. 
  Figure adapted from \citet{2011ApJS..194...28K}.}
\label{fig:reconstruct}
\end{figure}

\section{Reconstructing CV Evolution Empirically}

The observational tests described above provide strong
evidence that our basic ideas about CV evolution are at least
qualitatively correct. But does the standard model agree {\em
quantitatively} with observations? What is the strength of MB above
the period gap? Is GR really the only AML mechanism acting below the
gap? These issues are central not only to CVs, but to virtually all
types of close binaries, since  AML via MB and/or GR are thought to
drive the evolution of these systems also. 

Ideally, we would like to address such questions by reconstructing the
evolutionary path followed by CVs empirically. In practice, this means
that we want observations to tell us how the secular mass-transfer
rate in CVs depends on orbital period. The word ``secular'' is key
here, since it encapsulates the main difficulty in this project. The
problem is that most conventional tracers of $\dot{M}$ -- in
particular those tied to the accretion luminosity -- are
necessarily measures of the {\em instantaneous} mass transfer rate in
the system. However, from an evolutionary perspective, what we need is
the secular accretion rate, i.e. $\dot{M}$ averaged over evolutionary
time-scales. The trouble is that there is no guarantee that
instantaneous and long-term $\dot{M}$ are the same. In fact, it has
been known for a long time that CVs with apparently very different
instantaneous accretion rates (e.g. dwarf novae and nova-likes) can
co-exist at the same orbital periods. One possible explanation is that
CVs may undergo irradiation-driven mass-transfer cycles on time-scales
of $10^5$~yrs \citep[the thermal time-scale of the donor's envelope;
e.g.][]{2004A&A...423..281B}. 

Recent years have seen the emergence of {\em two} new methods to
overcome this problem. The first is based on the 
properties of the accreting WDs in CVs, the second on the properties
of their mass-losing donors. The WD-based method builds on the
theoretical work of Dean Townsley and Lars Bildsten, who have shown 
that the (quiescent) effective temperature of an accreting WD in a CV
is a tracer of $\dot{M}$ \citep{2002ApJ...565L..35T,2003ApJ...596L.227T}
The donor-based method, on the other hand, exploits the fact that CV
secondaries are driven out of thermal equilibrium, and hence inflated, by
mass loss (see Figure~\ref{fig:mr}). As discussed in detail in 
\citet{2011ApJS..194...28K}, this makes it possible to use the
degree of donor inflation as a tracer of secular
$\dot{M}$ \citep[see also][]{2010ApJ...721.1356S}. 

Both methods have their drawbacks, of course. WD-based $\dot{M}$
estimates are sensitive to the masses of the WD and its accreted
envelope (which are usually not well known), plus there remains 
a residual $T_{eff}$ response to long-term $\dot{M}$ variations, 
especially above the period gap. The main weaknesses of the
donor-based method are its strong reliance on theoretical models of
low-mass stars, as well as its sensitivity to 
apparent donor inflation unrelated to mass loss (e.g. due to 
tidal/rotatational deformation, or simply as a result of model
inadaquacies).

The first results obtained by the two methods are shown 
in Figure~\ref{fig:reconstruct}. The left panel is taken from 
\citet{2009ApJ...693.1007T} and shows how
$T_{eff}(P_{orb})$ predicted 
by different evolutionary models (including the standard one) compare
to a carefully compiled set of observed WD temperatures. The right
panel is adapted from \citet{2011ApJS..194...28K} and shows a similar
comparison between (standard and non-standard) models and data in the
donor mass-radius plane.  

A full discussion of these results would go far beyond the scope of
this review, so I will focus on just one important aspect. Taken
at face value, both methods seem to suggest that GR alone is not
sufficient to drive the observed mass-loss rates below the period
gap. However, much work remains to be done in testing these methods,
exploring their limitations, verifying such findings and studying
their implications. 
What is clear, however, is that we finally have the tools to
test the standard model quantitatively and, if necessary, to derive an 
empirically-calibrated alternative model that can be used as a
benchmark in population synthesis and other studies. In fact, the
best-fit donor-based model in the right panel of
Figure~\ref{fig:reconstruct} is intended to provide exactly such an
alternative \citep[for details, see][]{2011ApJS..194...28K}.

\section{Summary and Conclusions}

I hope this review has managed to get across that our understanding of
CVs and their evolution has improved dramatically over the last few
years. In particular, the long-standing fundamental predictions of
evolution theory are finally being tested observationally. We have
even learned to reconstruct CV evolution empirically, based on the
properties of the primary and secondary stars in these systems. 

Since the conference topic was on the topic of binary evolution, I
have not talked much at all about the actual accretion process in
CVs. However, here, too, there has been much progress in recent
years. In my view way, one of the most significant developments in
this area is the emerging recognition that virtually all facets of
this process -- including variability, disk winds and jets -- are
``universal'', with accreting WDs, neutron stars and black holes on
all scales exhibiting quantitatively similar phenemenology
\citep{2010AIPC.1314..171K}. In this context, CVs have the potential
to become prime laboratories for the underlying accretion physics.

\acknowledgements I am grateful to all of my many collaborators in the
CV community over the years. Joe Patterson, Brian Warner, 
Isabelle Baraffe, Boris G\"{a}nsicke, Stuart Littlefair, Joe
Patterson, Retha Pretorius and Brian Warner, in particular, have
strongly influenced my thinking about CV evolution. This article was
reprinted in part with permission from Knigge (2010; Copyright 2010,
American Institute of Physics) and from Knigge (2011); Copyright 2011,
Universal Academic Press).

\bibliographystyle{asp2010}
\bibliography{knigge}

\end{document}